\documentclass[aps, prd, twocolumn, superscriptaddress, showpacs, amsmath, amssymb, tightenlines]{revtex4-1}

\usepackage{graphicx} 
\usepackage{dcolumn}  
\usepackage{epstopdf}
\usepackage{bm}

\usepackage[absolute]{textpos}

\graphicspath{{ps}}

\newcommand{\bsig}{B_{\textrm{sig}}}
\newcommand{\btag}{B_{\textrm{tag}}}
\newcommand{\ecl}{E_{\textrm{ECL}}}
\newcommand{\psig}{p_{\textrm{sig}}^*}

\newcommand{\bdl}{\cos \theta_{B,D^{(*)}\ell}}
\newcommand{\tautoe}{\tau^+ \to e^+ \bar{\nu}_\tau \nu_e}
\newcommand{\tautomu}{\tau^+ \to \mu^+ \bar{\nu}_\tau \nu_\mu}
\newcommand{\tautopi}{\tau^+ \to \pi^+ \bar{\nu}_\tau}
\newcommand{\tautorho}{\tau^+ \to \rho^+ \bar{\nu}_\tau}
\newcommand{\Bdecay}{B^+ \to \tau^+ \nu_\tau}
\newcommand{\bbar}{B \bar{B}}

\newcommand{\mevcc}{{\mathrm{\,Me\kern -0.1em V\!/}c^2}}
\newcommand{\mevc}{{\mathrm{\,Me\kern -0.1em V\!/}c}}
\newcommand{\gevcc}{{\mathrm{\,Ge\kern -0.1em V\!/}c^2}}
\newcommand{\gevc}{{\mathrm{\,Ge\kern -0.1em V\!/}c}}

\begin{document}

\hbox{Belle-CONF-1401}

\title{\quad\\[1.0cm] \boldmath Measurement of the branching fraction of $\Bdecay$ decays with the semileptonic tagging method and the full Belle data sample}

\noaffiliation
\affiliation{University of the Basque Country UPV/EHU, 48080 Bilbao}
\affiliation{Beihang University, Beijing 100191}
\affiliation{University of Bonn, 53115 Bonn}
\affiliation{Budker Institute of Nuclear Physics SB RAS and Novosibirsk State University, Novosibirsk 630090}
\affiliation{Faculty of Mathematics and Physics, Charles University, 121 16 Prague}
\affiliation{Chiba University, Chiba 263-8522}
\affiliation{Chonnam National University, Kwangju 660-701}
\affiliation{University of Cincinnati, Cincinnati, Ohio 45221}
\affiliation{Deutsches Elektronen--Synchrotron, 22607 Hamburg}
\affiliation{Department of Physics, Fu Jen Catholic University, Taipei 24205}
\affiliation{Justus-Liebig-Universit\"at Gie\ss{}en, 35392 Gie\ss{}en}
\affiliation{Gifu University, Gifu 501-1193}
\affiliation{II. Physikalisches Institut, Georg-August-Universit\"at G\"ottingen, 37073 G\"ottingen}
\affiliation{The Graduate University for Advanced Studies, Hayama 240-0193}
\affiliation{Gyeongsang National University, Chinju 660-701}
\affiliation{Hanyang University, Seoul 133-791}
\affiliation{University of Hawaii, Honolulu, Hawaii 96822}
\affiliation{High Energy Accelerator Research Organization (KEK), Tsukuba 305-0801}
\affiliation{Hiroshima Institute of Technology, Hiroshima 731-5193}
\affiliation{IKERBASQUE, Basque Foundation for Science, 48011 Bilbao}
\affiliation{University of Illinois at Urbana-Champaign, Urbana, Illinois 61801}
\affiliation{Indian Institute of Technology Bhubaneswar, Satya Nagar 751007}
\affiliation{Indian Institute of Technology Guwahati, Assam 781039}
\affiliation{Indian Institute of Technology Madras, Chennai 600036}
\affiliation{Indiana University, Bloomington, Indiana 47408}
\affiliation{Institute of High Energy Physics, Chinese Academy of Sciences, Beijing 100049}
\affiliation{Institute of High Energy Physics, Vienna 1050}
\affiliation{Institute for High Energy Physics, Protvino 142281}
\affiliation{Institute of Mathematical Sciences, Chennai 600113}
\affiliation{INFN - Sezione di Torino, 10125 Torino}
\affiliation{Institute for Theoretical and Experimental Physics, Moscow 117218}
\affiliation{J. Stefan Institute, 1000 Ljubljana}
\affiliation{Kanagawa University, Yokohama 221-8686}
\affiliation{Institut f\"ur Experimentelle Kernphysik, Karlsruher Institut f\"ur Technologie, 76131 Karlsruhe}
\affiliation{Kavli Institute for the Physics and Mathematics of the Universe (WPI), University of Tokyo, Kashiwa 277-8583}
\affiliation{Department of Physics, Faculty of Science, King Abdulaziz University, Jeddah 21589}
\affiliation{Korea Institute of Science and Technology Information, Daejeon 305-806}
\affiliation{Korea University, Seoul 136-713}
\affiliation{Kyoto University, Kyoto 606-8502}
\affiliation{Kyungpook National University, Daegu 702-701}
\affiliation{\'Ecole Polytechnique F\'ed\'erale de Lausanne (EPFL), Lausanne 1015}
\affiliation{Faculty of Mathematics and Physics, University of Ljubljana, 1000 Ljubljana}
\affiliation{Luther College, Decorah, Iowa 52101}
\affiliation{University of Maribor, 2000 Maribor}
\affiliation{Max-Planck-Institut f\"ur Physik, 80805 M\"unchen}
\affiliation{School of Physics, University of Melbourne, Victoria 3010}
\affiliation{Moscow Physical Engineering Institute, Moscow 115409}
\affiliation{Moscow Institute of Physics and Technology, Moscow Region 141700}
\affiliation{Graduate School of Science, Nagoya University, Nagoya 464-8602}
\affiliation{Kobayashi-Maskawa Institute, Nagoya University, Nagoya 464-8602}
\affiliation{Nara University of Education, Nara 630-8528}
\affiliation{Nara Women's University, Nara 630-8506}
\affiliation{National Central University, Chung-li 32054}
\affiliation{National United University, Miao Li 36003}
\affiliation{Department of Physics, National Taiwan University, Taipei 10617}
\affiliation{H. Niewodniczanski Institute of Nuclear Physics, Krakow 31-342}
\affiliation{Nippon Dental University, Niigata 951-8580}
\affiliation{Niigata University, Niigata 950-2181}
\affiliation{University of Nova Gorica, 5000 Nova Gorica}
\affiliation{Osaka City University, Osaka 558-8585}
\affiliation{Osaka University, Osaka 565-0871}
\affiliation{Pacific Northwest National Laboratory, Richland, Washington 99352}
\affiliation{Panjab University, Chandigarh 160014}
\affiliation{Peking University, Beijing 100871}
\affiliation{University of Pittsburgh, Pittsburgh, Pennsylvania 15260}
\affiliation{Punjab Agricultural University, Ludhiana 141004}
\affiliation{Research Center for Electron Photon Science, Tohoku University, Sendai 980-8578}
\affiliation{Research Center for Nuclear Physics, Osaka University, Osaka 567-0047}
\affiliation{RIKEN BNL Research Center, Upton, New York 11973}
\affiliation{Saga University, Saga 840-8502}
\affiliation{University of Science and Technology of China, Hefei 230026}
\affiliation{Seoul National University, Seoul 151-742}
\affiliation{Shinshu University, Nagano 390-8621}
\affiliation{Soongsil University, Seoul 156-743}
\affiliation{Sungkyunkwan University, Suwon 440-746}
\affiliation{School of Physics, University of Sydney, NSW 2006}
\affiliation{Department of Physics, Faculty of Science, University of Tabuk, Tabuk 71451}
\affiliation{Tata Institute of Fundamental Research, Mumbai 400005}
\affiliation{Excellence Cluster Universe, Technische Universit\"at M\"unchen, 85748 Garching}
\affiliation{Toho University, Funabashi 274-8510}
\affiliation{Tohoku Gakuin University, Tagajo 985-8537}
\affiliation{Tohoku University, Sendai 980-8578}
\affiliation{Department of Physics, University of Tokyo, Tokyo 113-0033}
\affiliation{Tokyo Institute of Technology, Tokyo 152-8550}
\affiliation{Tokyo Metropolitan University, Tokyo 192-0397}
\affiliation{Tokyo University of Agriculture and Technology, Tokyo 184-8588}
\affiliation{University of Torino, 10124 Torino}
\affiliation{Toyama National College of Maritime Technology, Toyama 933-0293}
\affiliation{CNP, Virginia Polytechnic Institute and State University, Blacksburg, Virginia 24061}
\affiliation{Wayne State University, Detroit, Michigan 48202}
\affiliation{Yamagata University, Yamagata 990-8560}
\affiliation{Yonsei University, Seoul 120-749}
  \author{A.~Abdesselam}\affiliation{Department of Physics, Faculty of Science, University of Tabuk, Tabuk 71451} 
  \author{I.~Adachi}\affiliation{High Energy Accelerator Research Organization (KEK), Tsukuba 305-0801}\affiliation{The Graduate University for Advanced Studies, Hayama 240-0193} 
  \author{K.~Adamczyk}\affiliation{H. Niewodniczanski Institute of Nuclear Physics, Krakow 31-342} 
  \author{H.~Aihara}\affiliation{Department of Physics, University of Tokyo, Tokyo 113-0033} 
  \author{S.~Al~Said}\affiliation{Department of Physics, Faculty of Science, University of Tabuk, Tabuk 71451}\affiliation{Department of Physics, Faculty of Science, King Abdulaziz University, Jeddah 21589} 
  \author{K.~Arinstein}\affiliation{Budker Institute of Nuclear Physics SB RAS and Novosibirsk State University, Novosibirsk 630090} 
  \author{Y.~Arita}\affiliation{Graduate School of Science, Nagoya University, Nagoya 464-8602} 
  \author{D.~M.~Asner}\affiliation{Pacific Northwest National Laboratory, Richland, Washington 99352} 
  \author{T.~Aso}\affiliation{Toyama National College of Maritime Technology, Toyama 933-0293} 
  \author{V.~Aulchenko}\affiliation{Budker Institute of Nuclear Physics SB RAS and Novosibirsk State University, Novosibirsk 630090} 
% ...
\collaboration{The Belle Collaboration}

\begin{abstract}
We report a measurement of the branching fraction of $\Bdecay$ decays using a data sample of $772 \times 10^6 \bbar$ pairs, collected at the $\Upsilon(4S)$ resonance with the Belle detector at the KEKB asymmetric-energy $e^+e^-$ collider. We reconstruct the accompanying $B$ meson in a semileptonic decay and detect the $\Bdecay$ candidate in the recoiling event.
We obtain a branching fraction of ${\cal B}(\Bdecay) = [1.25 \pm 0.28 ({\rm stat.}) \pm 0.27({\rm syst.})] \times 10^{-4}$. This result is in good agreement with previous measurements and the expectation from calculations based on the Standard Model.
\end{abstract}


\maketitle

{\renewcommand{\thefootnote}{\fnsymbol{footnote}}}
\setcounter{footnote}{0}
In the Standard Model (SM) the branching fraction of the purely leptonic decay $\Bdecay$\cite{CC} is given by
\begin{equation}
 {\cal B}(\Bdecay)_{\rm SM} = \frac{G_F^2 m_B m_\tau^2}{8 \pi} \left( 1 - \frac{m_\tau^2}{m_B^2} \right)^2 f_B^2 |V_{ub}|^2 \tau_B, \label{equ:BR}
\end{equation}
where $G_F$ is the Fermi coupling constant, $V_{ub}$ the Cabibbo-Kobayashi-Maskawa matrix element, $m_B$ and $m_\tau$ the masses of the $B$ meson and the $\tau$ lepton, respectively, $\tau_B$ the lifetime of the $B$ meson, and $f_B$ the $B$-meson decay constant.
The branching fraction dependends on the mass of the lepton strongly by the factor $m_\tau^2$ because of the helicity suppression and weakly by the phase space factor $(1 - {m_\tau^2}/{m_B^2})^2$. Therefore $\Bdecay$ is expected to be the highest purely leptonic branching fraction of the $B^+$ meson and is the only decay of this kind which has been measured with a significance of more than three standard deviations.
All of the inputs of Eq.~\ref{equ:BR} are measured or, in the case of $f_B$, can be obtained using the methods of lattice quantum chromodynamics. An independent estimation of the branching fraction, which uses $V_{ub} = (3.70 \pm 0.12 \pm 0.26) \times 10^{-3}$, $f_{B_s} = (225.6 \pm 1.1 \pm 5.4)$ MeV, and $f_{B_s} / f_{B_d} = 1.205 \pm 0.004 \pm 0.007$ as input, gives ${\cal B}(\Bdecay) = (0.753^{+0.102}_{-0.052}) \times 10^{-4}$~\cite{CKMfitter}.

\par
Physics beyond the SM, such as the presence of additional charged Higgs bosons, could constructively or destructively interfere with the SM weak decay process. Measurements by the BaBar~\cite{BaBarLep,BaBarHad} and Belle~\cite{BelleLep} collaborations showed a slight disagreement with the SM expectation, but the most recent measurement by the Belle collaboration~\cite{BelleHad}, using a hadronic tagging method, was in very good agreement. The current world average ${\cal B}(\Bdecay) = (1.14 \pm 0.27) \times 10^{-4}$~\cite{PDG} shows no sign of physics beyond the SM. This average is obtained inflating the uncertainties of the input values by a factor of $1.22$ to take into account discrepancies between the recent measurements.

\par
The measurement described in this paper is performed using the final Belle data sample consisting of an integrated luminosity of $711 \,{\rm fb}^{-1}$ containing $(772 \pm 11) \times 10^6 \bbar$ pairs, collected at the $\Upsilon(4S)$ resonance at the KEKB asymmetric-energy $e^+e^-$ collider~\cite{kekb}. We also use a smaller data sample with an integrated luminosity of $79 \,{\rm fb}^{-1}$ taken at an energy lower than the $\Upsilon(4S)$ mass to study the background from continuum $e^+ e^- \to q \bar{q} \, (q = u, d, s, c)$ events and other processes without $b$-quark production.
We generate multiple samples of simulated Monte Carlo (MC) events. We first simulate the decays to the final state using the software package EvtGen~\cite{evtgen}, and then the interaction with the detector and its response using GEANT3~\cite{GEANT}. The simulated signal events are overlaid by beam related background, which was recorded with a random trigger.
 
\par 
The Belle detector is a large-solid-angle magnetic spectrometer that consists of a silicon vertex detector, a 50-layer central drift chamber (CDC), an array of aerogel threshold Cherenkov counters (ACC), a barrel-like arrangement of time-of-flight scintillation counters (TOF), and an electromagnetic calorimeter composed of CsI(Tl) crystals (ECL) located inside a superconducting solenoid coil that provides a 1.5~T magnetic field. An iron flux-return located outside the coil is instrumented to detect $K_L^0$ mesons and to identify muons (KLM). The detector is described in detail elsewhere~\cite{Belle}.
Two inner detector configurations were used. A 2.0 cm beam-pipe and a 3-layer silicon vertex detector were used for the first sample of $152 \times 10^6 B\bar{B}$ pairs, while a 1.5 cm beam-pipe, a 4-layer silicon detector and a small-cell inner drift chamber were used to record the remaining $620 \times 10^6 B\bar{B}$ pairs~\cite{svd2}.

\par 
Since the detectable signature of a $\Bdecay$ decay is often only a single charged track, we reconstruct the accompanying $B$ meson (referred to as $\btag$) in the semileptonic decay channels $B^+ \to D^{*0} \ell^+ \nu_\ell$ and $B^+ \to D^{0} \ell^+ \nu_\ell$, where $\ell$ can be an electron or muon. The $D^{*0}$ mesons are reconstructed as $D^{*0} \to D^0 \pi^0$ and $D^{*0} \to D^0 \gamma$ and the $D^0$ mesons as $D^0 \to K^- \pi^+ \pi^0$, $K^- \pi^+ \pi^+ \pi^-$, $K^0_S \pi^+ \pi^- \pi^0$, $K^- \pi^+$, $K^0_S \pi^+ \pi^-$, $\pi^+ \pi^- \pi^0$, $K^0_S \pi^0$, $K^0_S K^+ K^-$, $K^+ K^-$, and $\pi^+ \pi^-$. Neutral pions are reconstructed as $\pi^0 \to \gamma \gamma$ and $K_S^0$ as $K_S^0 \to \pi^+ \pi^-$.
\par
To maximize the efficiency in reconstructing $\btag$ candidates, only loose requirements are applied. Charged final state particles are selected from well-measured tracks and are required to have a distance to the interaction point along (perpendicular to) the beam direction, further denoted as $dz (dr)$, of less than $4 ~ (2)$~cm. Photons used for the reconstruction of neutral pions are required to have an energy of at least $30$ MeV and the invariant mass of the two-photon system must satisfy $|M_{\gamma \gamma} - m_{\pi^0}| < 19 \mevcc$; this corresponds to a width of $3.2 \sigma$. The invariant mass of the two charged tracks which are used to form $K_S^0$ candidates must lie within $30 \mevcc (4.5 \sigma)$ of the nominal $K_S^0$ mass. The momenta of $D^{(*)0}$ meson candidates are required to be below $2.5 \gevc$. All further selection is performed by training a multivariate selection (MVS) method, based on the NeuroBayes package~\cite{NB}. A large sample of generically decaying simulated $B$ mesons is used for this training and a broad range of information is considered in each stage of the reconstruction. Commonly used information is the mass, momentum, and decay channel of the particle candidate, as well as momenta, angles, and the output of the MVS of daughter particles. The structure of this semileptonic reconstruction method is very similar to the existing hadronic full reconstruction method~\cite{HadFR}. The variables were chosen to be uncorrelated to the cosine of the angle between the momentum of the $B$ meson and the $D^{(*)}\ell$ system, calculated under the assumption that only one massless particle is not reconstructed. It is given by 
\begin{equation}
\bdl = \frac{2 E_{\rm beam} E_{D^{(*)} \ell} - m_B^2 - m^2_{D^{(*)} \ell}}{2 p^*_B p^*_{D^{(*)} \ell}}, \label{equ:bdl}
\end{equation}
where $E_{\rm beam}$ is the energy of the beam in the center-of-mass system (CMS), $E_{D^{(*)} \ell}$, $m^2_{D^{(*)} \ell}$ and $p^*_{D^{(*)} \ell}$ are the energy, mass and momentum of the $(D^{(*)} \ell)$ system in the CMS, respectively, $m_B$ is the nominal $B$ meson mass~\cite{PDG}, and $p^*_B$ is the nominal $B$ meson momentum in the CMS, calculated from the beam energy and the nominal mass. This angle is used later for further selection, since correctly reconstructed $\btag$ candidates have values between $-1$ and $1$, while background events, where the assumption of only one missing massless particle does not hold, have a much larger range of values. Partially reconstructed $\btag$ candidates where only the slow pion or soft photon is not reconstructed lie in a broader range, but still peak around the signal region.

\par
The $\btag$ candidates are combined with $B$ mesons reconstructed in the decay mode $\Bdecay$, further denoted as $\bsig$. The $\tau$ lepton is reconstructed as $\tautomu$, $e^+ \bar{\nu}_\tau \nu_e$, $\pi^+ \bar{\nu}_\tau$, and $\rho^+ \bar{\nu}_\tau$, where the $\rho^+$ is reconstructed as $\rho^+ \to \pi^+ \pi^0$.  
Since the neutrinos cannot be detected, the $\bsig$ candidate consists only of a single charged track or a $\rho^+$ candidate. The $\rho^+$ candidate is required to have an invariant mass within of $195 \mevcc$ of the nominal $\rho^+$ mass.
The signal side particles are separated based on particle identification variables. The pion and kaon separation uses information from the ACC, TOF, and the $dE/dx$ measurement in the CDC; the electron identification is based on the same information in addition to the shape of the shower and the energy measurement in the ECL; and muon candidates are identified using hits in the KLM matched to a charged track. The selection is performed such that signal side particle(s) can only be reconstructed in one of the potential particle hypotheses. The momentum of the signal side particle ($e^+, \mu^+, \pi^+$, or $\rho^+$) in the CMS ($\psig$) must be in the range \mbox{$0.5 \gevc < \psig < 2.4 \gevc$.} 

\par
The combination of a $\btag$ and a $\bsig$ candidate is identical to the reconstruction of the $\Upsilon (4S)$. Since the $\Upsilon (4S)$ is produced without any accompanying particles, this allows for a powerful form of selection: we therefore reject events with additional $\pi^0$ candidates or charged tracks with $|dz| < 100$ cm and $|dr| < 20$ cm.
In the decay channel $\tautoe$, a significant background arises from events containing converted photons. To suppress this, we combine the electron, either the one used in the reconstruction of the $\btag$ candidate or the one in the signal side, with every other oppositely charged track in the event. Using the electron mass hypothesis for the unspecific track, we require the invariant mass of the electron-track pair to be greater than $200 \mevcc$ for any of the pairs.
To suppress background from continuum events, we train another MVS with the following input variables: the polar angle of the $\btag$ candidate with respect to the beam direction in the CMS; the polar angle between the thrust axis of the $\btag$ candidate and the remaining tracks in the event in the CMS; 16 modified Fox–Wolfram moments~\cite{FWM}; and the momentum flow in nine concentric cones around the thrust axis of the $\btag$ candidate~\cite{CleoCones}. The requirement on the output of the MVS depends on the $\tau$ decay channel since the continuum background contribution differs significantly between them.
The selection on $\bdl$ also differs between the $\tau$ decay channels. It is required to be smaller than $1$ in all channels, but the lower limits are $-1.7$, $-1.9$, $-1.3$, and $-2.6$ for the muon, electron, pion, and $\rho$ final state, respectively.
The selection is optimized using samples of simulated signal and background events to give the highest Figure of Merit $N_S / \sqrt{N_S + N_B}$, where $N_S$ and $N_B$ are the number of selected signal and background events, respectively.

\par
We additionally perform a selection in the remaining energy in the ECL, further denoted as $\ecl$. It is defined as the sum of the energies of clusters in the ECL that are not associated to a final state particle of the reconstructed $\Upsilon (4S)$ candidate. To mitigate beam induced background in the energy sum, only clusters satisfying minimum energy thresholds of $50$, $100$, and $150$ MeV are required for the barrel, forward, and backward end-cap calorimeter, respectively. Signal events peak near low values of $\ecl$ as only photons from beam related background and misreconstructed events contribute, while the background is distributed over a much wider range. We require $\ecl$ to be smaller than $1.2$ GeV.
The fraction of events with multiple signal candidates is $7\%$. In events with multiple candidates we choose the candidate with a maximal value of the tag side MVS classifier output. From MC simulation we find that this method selects the best candidate $70\%$ of the time.
The selection gives a total reconstruction efficiency of $\epsilon = (23.1 \pm 0.1) \times 10^{-4}$, where the uncertainty is due to MC statistics only. It is described in detail in Table~\ref{tab:eff}.

\begin{table}[htb]
\begin{tabular}{l c @{\hspace{1em}}c@{\hspace{1em}} c@{\hspace{1em}} c} 
\hline 
\hline 
        Final State & $e^+ \nu_e \bar{\nu}_\tau$         & $\mu^+ \nu_\mu \bar{\nu}_\tau$       & $\pi^+ \bar{\nu}_\tau$       & $\pi^+ \pi^0 \bar{\nu}_\tau$ \\ 
\hline 
$e^+ \nu_e \bar{\nu}_\tau$         & $6.6 \pm 0.1$ & $0.1 \pm 0.0$ & $0.2 \pm 0.0$ & $0.1 \pm 0.0$\\ 
$\mu^+ \nu_\mu \bar{\nu}_\tau$     & $0.1 \pm 0.0$ & $4.7 \pm 0.1$ & $0.6 \pm 0.0$ & $0.2 \pm 0.0$\\ 
$\pi^+ \bar{\nu}_\tau$             & $0$           & $0.1 \pm 0.0$ & $1.6 \pm 0.0$ & $0.5 \pm 0.0$\\
$\pi^+ \pi^0 \bar{\nu}_\tau$       & $0$           & $0.1 \pm 0.0$ & $1.4 \pm 0.0$ & $4.9 \pm 0.1$\\
$\pi^+ \pi^0 \pi^0 \bar{\nu}_\tau$ & $0$           & $0$           & $0.2 \pm 0.0$ & $1.3 \pm 0.0$\\ 
Other                              & $0$           & $0$           & $0.1 \pm 0.0$ & $0.2 \pm 0.0$\\ 
All                                & $6.8 \pm 0.1$ & $5.1 \pm 0.1$ & $4.0 \pm 0.0$ & $7.2 \pm 0.1$\\
Total               & \multicolumn{4}{c}{$23.1 \pm 0.1$} \\
                                        
\hline 
\hline 
\end{tabular}
\caption{Reconstruction efficiency $(\times 10^{-4})$ for each $\tau$ decay mode, determined from MC and corrected according to control sample studies. The row denotes the generated decay mode, and the columns represent the reconstructed final state. The off-diagonal entries reflect mode cross-feed.}
\label{tab:eff}
\end{table}

\par
To study possible differences between real and simulated data, we use samples where the $\bsig$ is reconstructed in the decays $B^+ \to D^{*0} \ell^+ \nu_\ell$ and $B^+ \to D^0 \pi^+$ (further denoted as double-tagged samples). The $D^{*0}$ mesons are reconstructed as $D^{*0} \to D^0 \pi^0$ and $D^{*0} \to D^0 \gamma$ and the $D^0$ meson as $D^0 \to K^- \pi^+$. The $D^*$ and $D$ meson candidates are selected based on their mass and the mass difference between the $D^*$ and the $D$ meson candidate. All selection related to the $\btag$ and the event-wide vetoes are applied in addition to the signal side selection. There is a set of selection criteria for each of the $\tau$ decay channels. We apply each of these sets of selection criteria on each of the double-tagged samples, thus produce four samples for every $B$ decay channel, which only differ in the tag-side related selection. We measure the branching fractions of the $B$ decays and compare them to the current world averages~\cite{PDG}. The reconstruction efficiency is corrected based on this ratio, depending on the decay channel of the $\btag$ and the $\tau$. The reconstruction efficiency is found to be overestimated by a factor of $1.02$ to $1.18$ in MC simulation.

\par
\begin{figure}[th]
 \includegraphics[width=.23\textwidth]{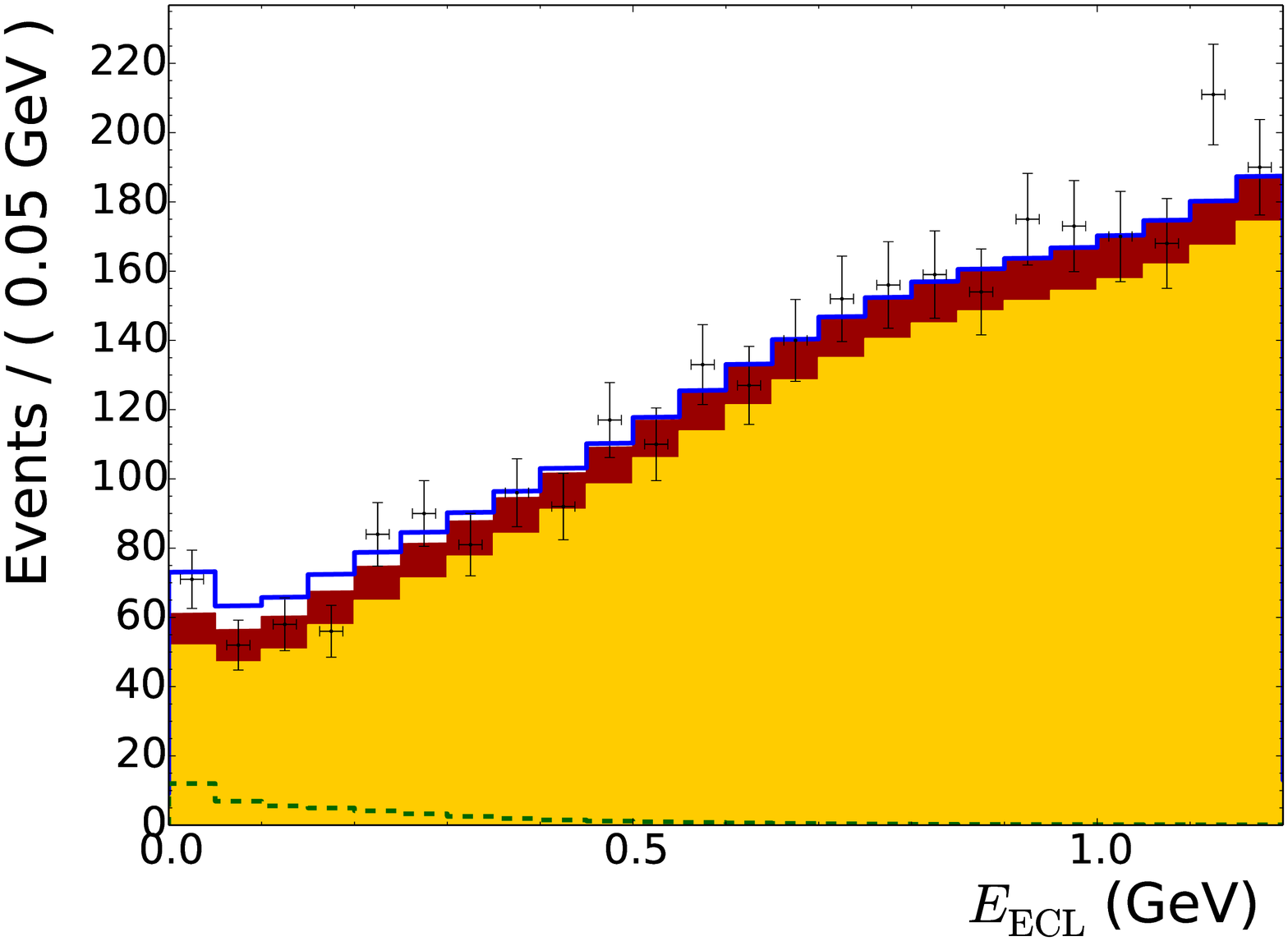} \put(-90,60){(a)}
 \includegraphics[width=.23\textwidth]{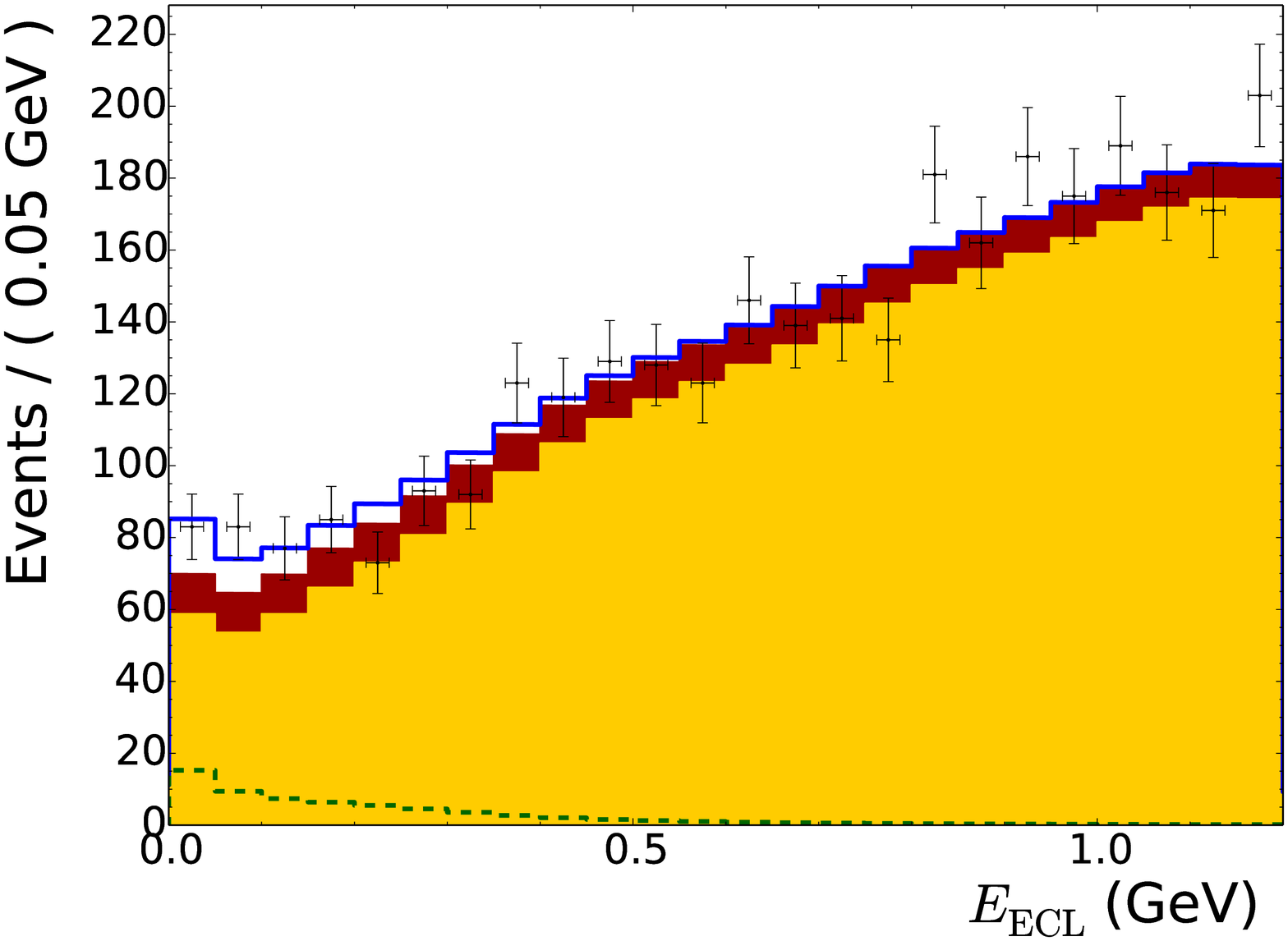} \put(-90,60){(b)} \\
 \includegraphics[width=.23\textwidth]{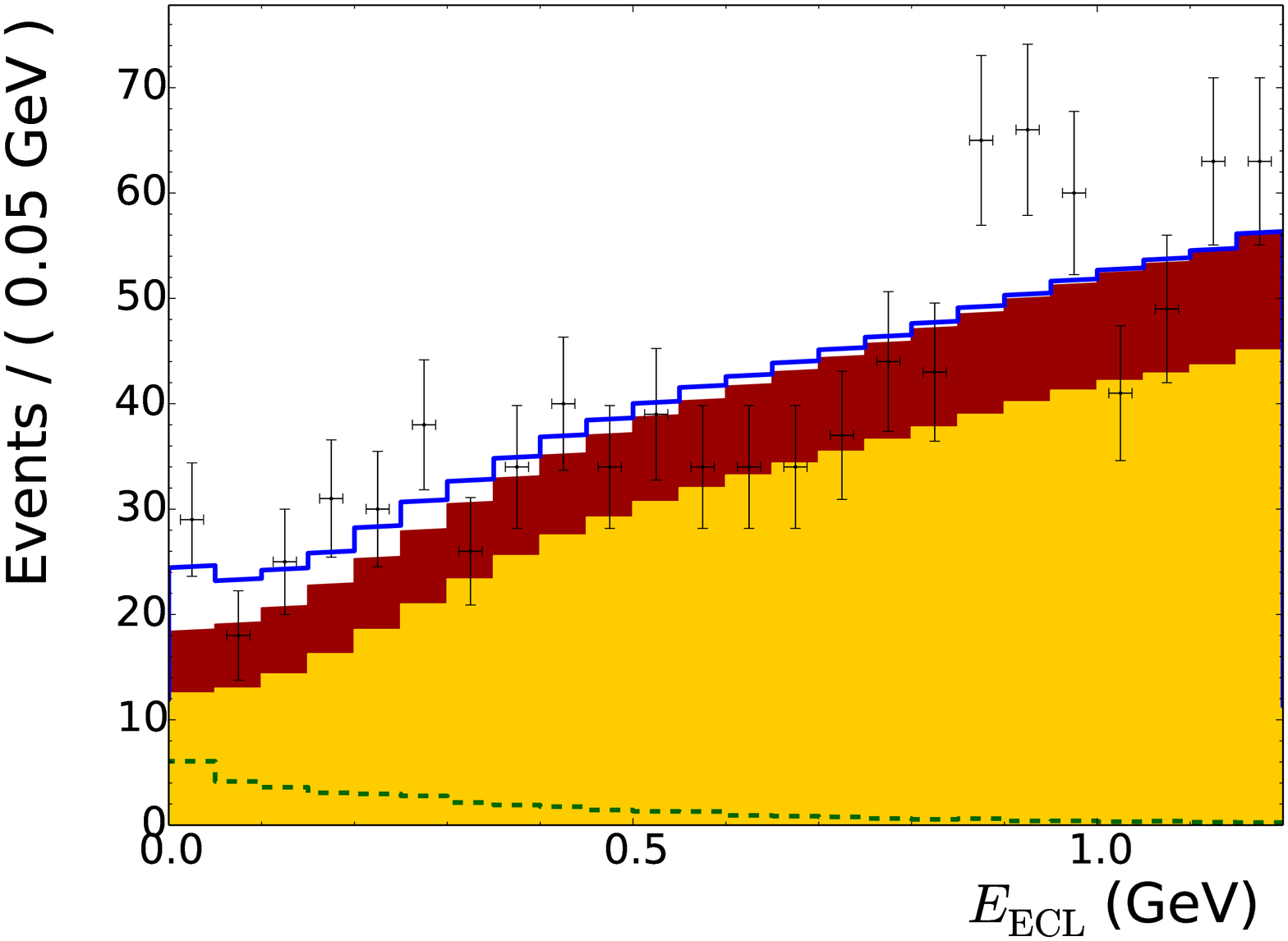} \put(-90,60){(c)}
 \includegraphics[width=.23\textwidth]{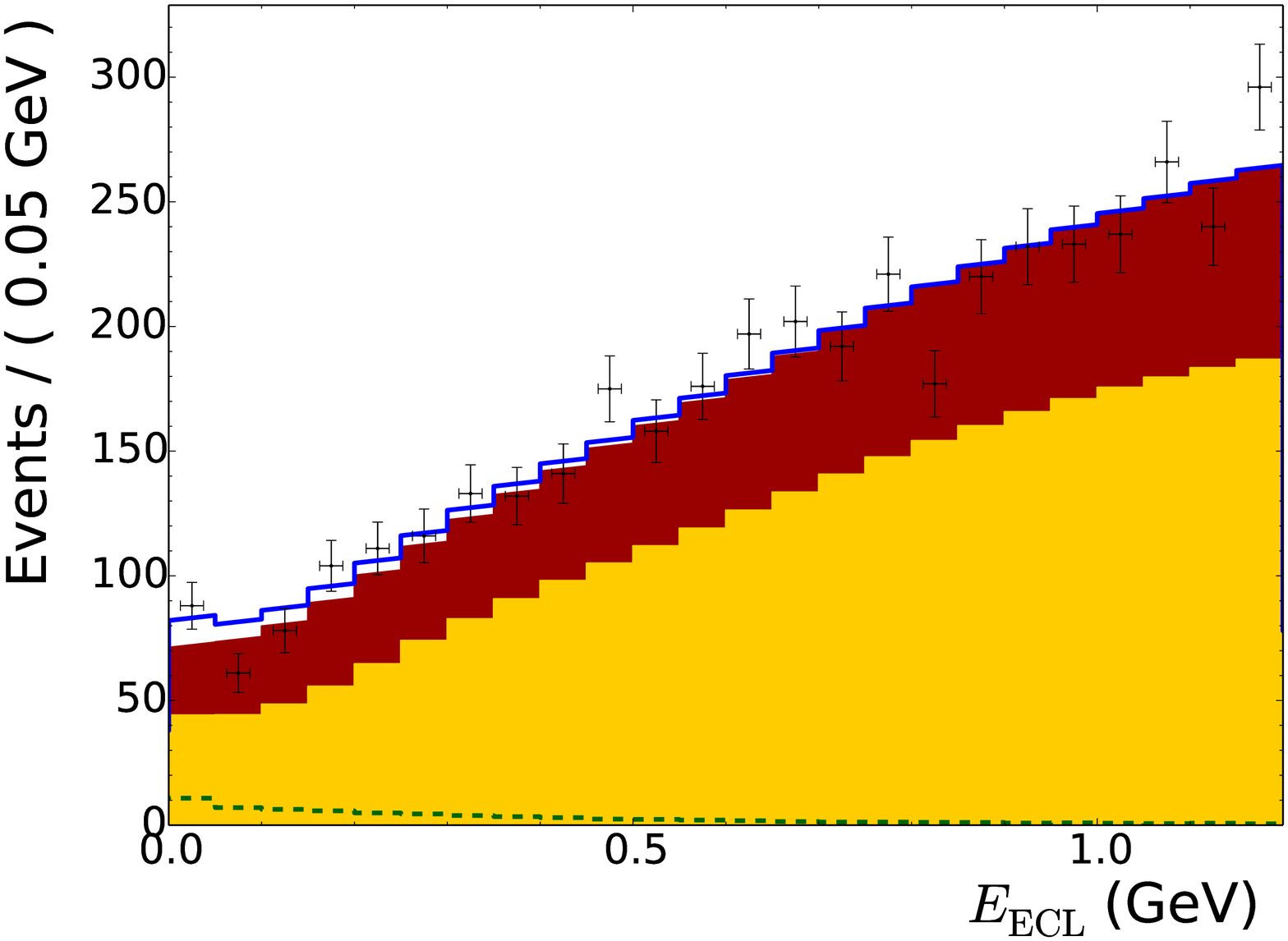} \put(-90,60){(d)}\\
\caption{Distribution of $\ecl$ for (a)~$\tautomu$, (b)~$\tautoe$, (c)~$\tautopi$, and (d)~$\tautorho$. The markers show the data distribution, the solid line the total fitted distribution, and the dashed line the signal component. The orange (red) filled distribution represents the $B \bar{B}$ (continuum) background.}
\label{Fig:ecl}
 \end{figure}
\begin{figure}[th]
 \includegraphics[width=.23\textwidth]{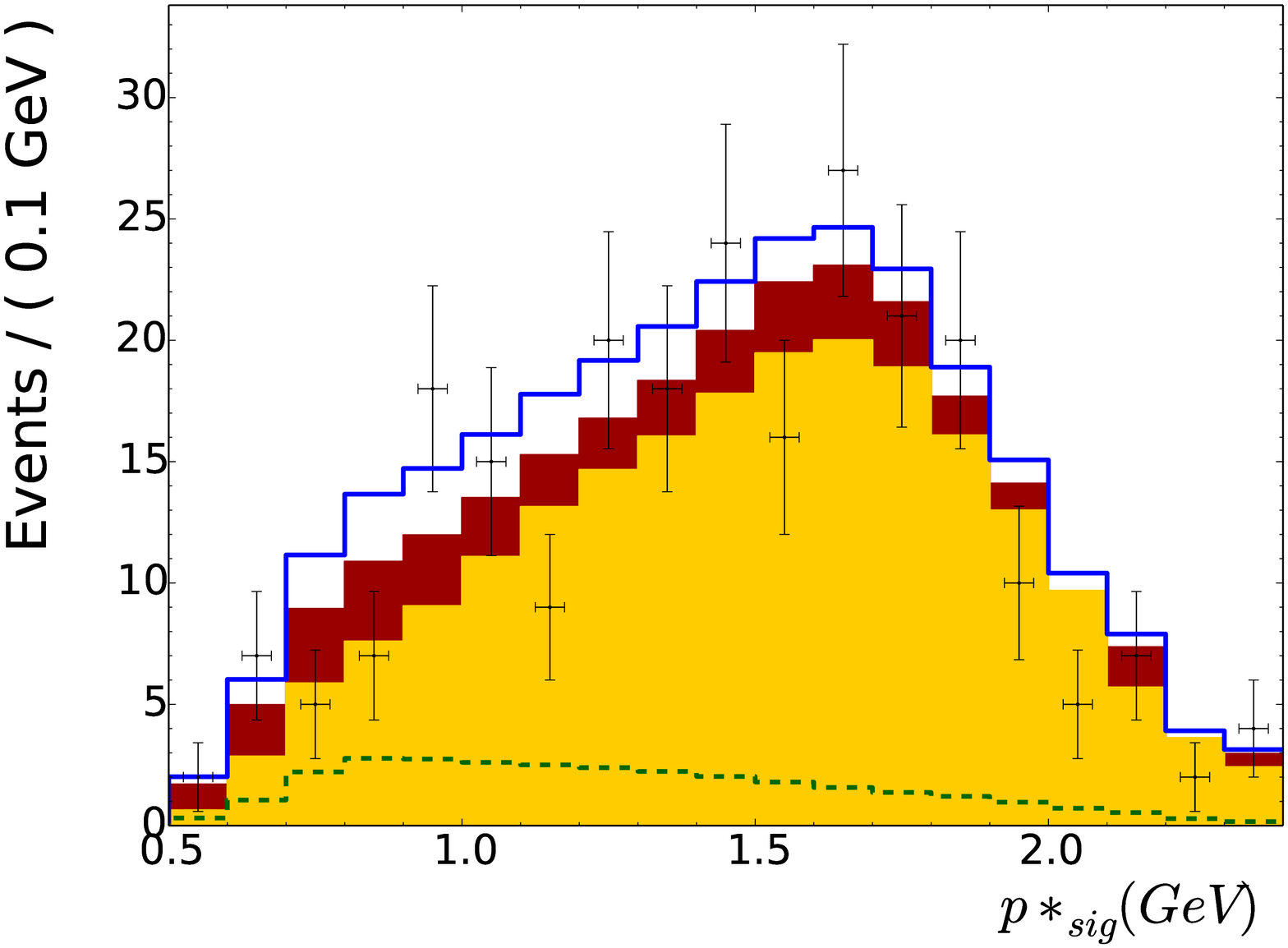} \put(-30,60){(a)}
 \includegraphics[width=.23\textwidth]{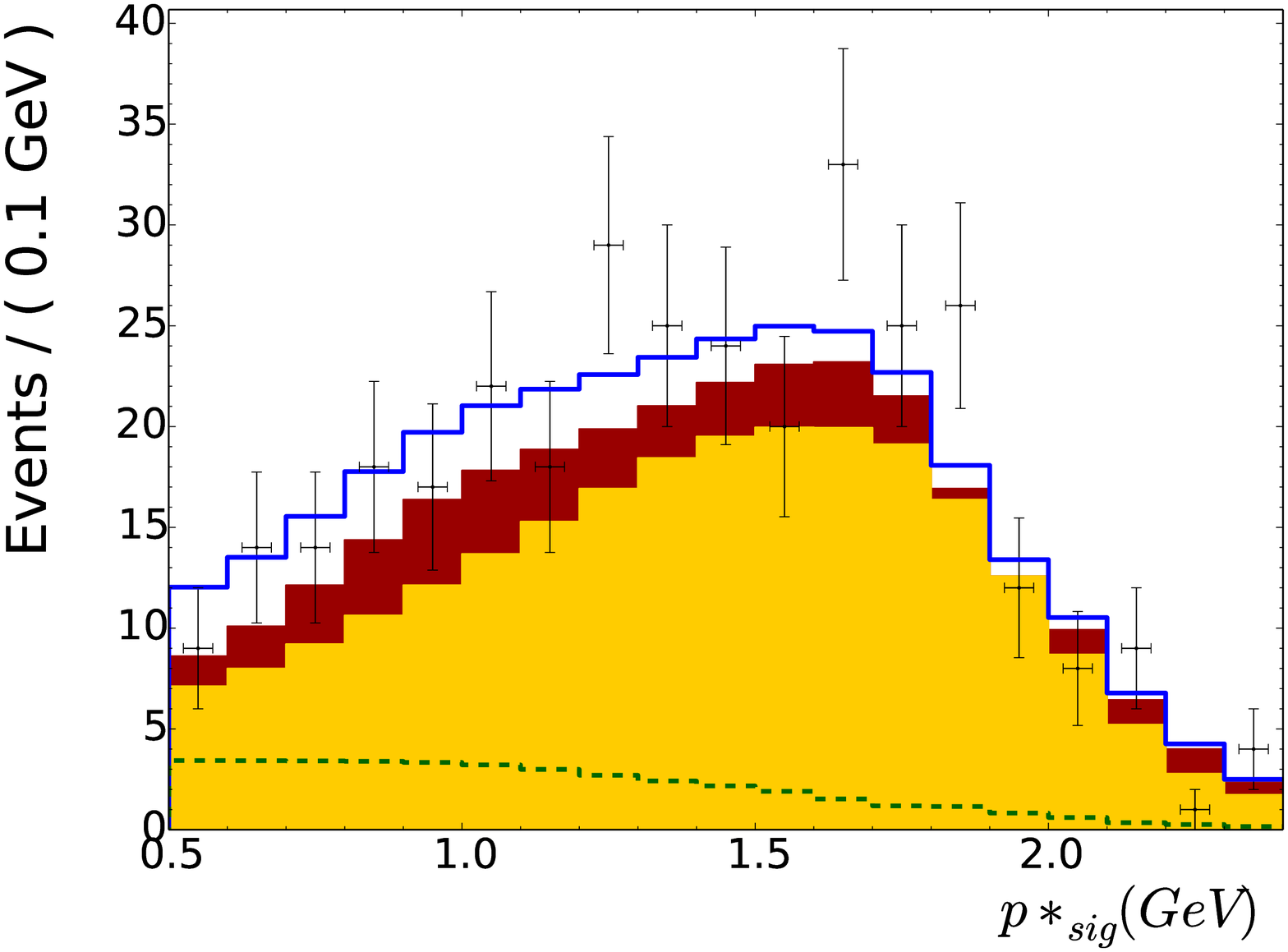} \put(-30,60){(b)} \\
 \includegraphics[width=.23\textwidth]{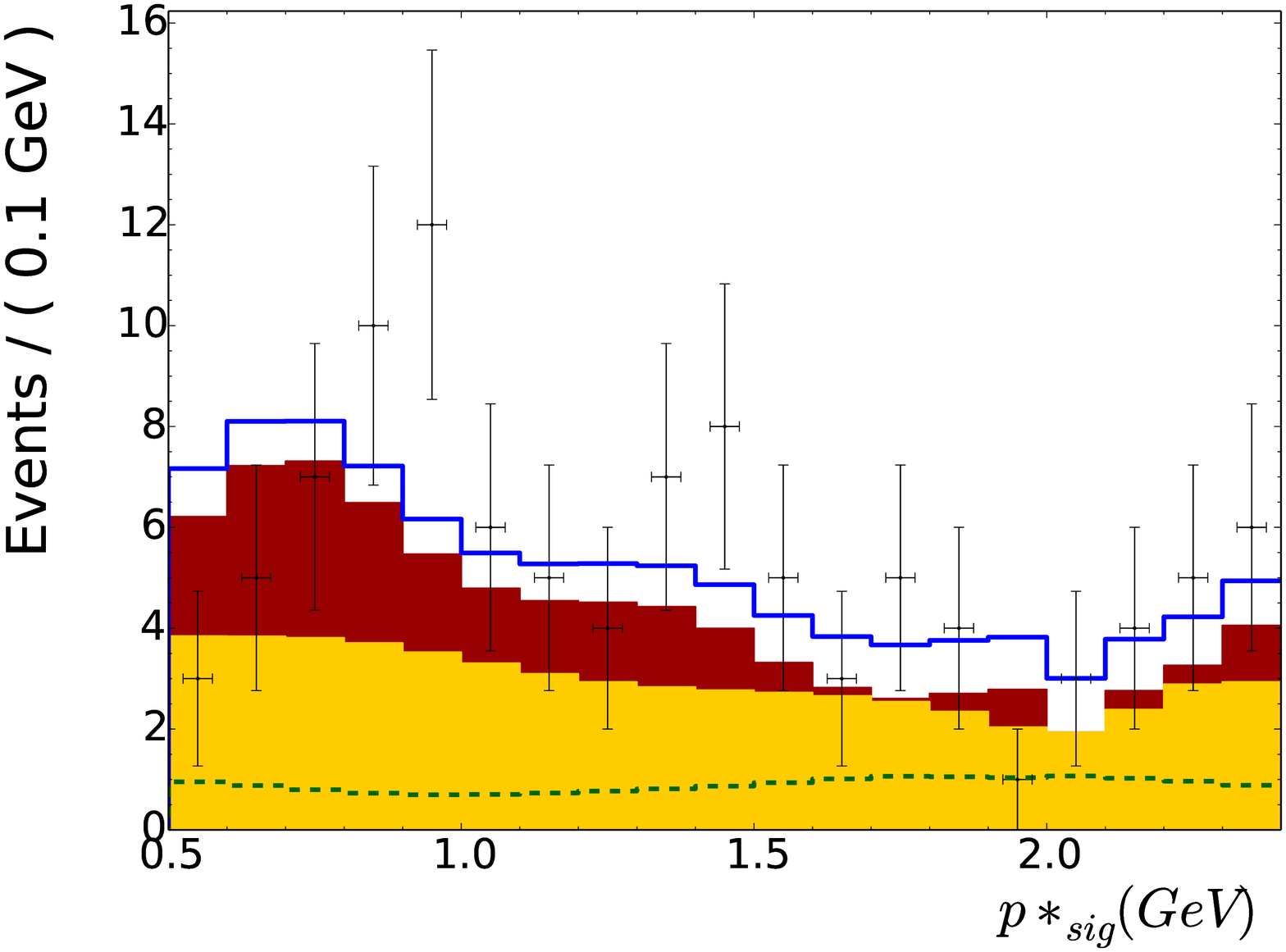} \put(-30,60){(c)}
 \includegraphics[width=.23\textwidth]{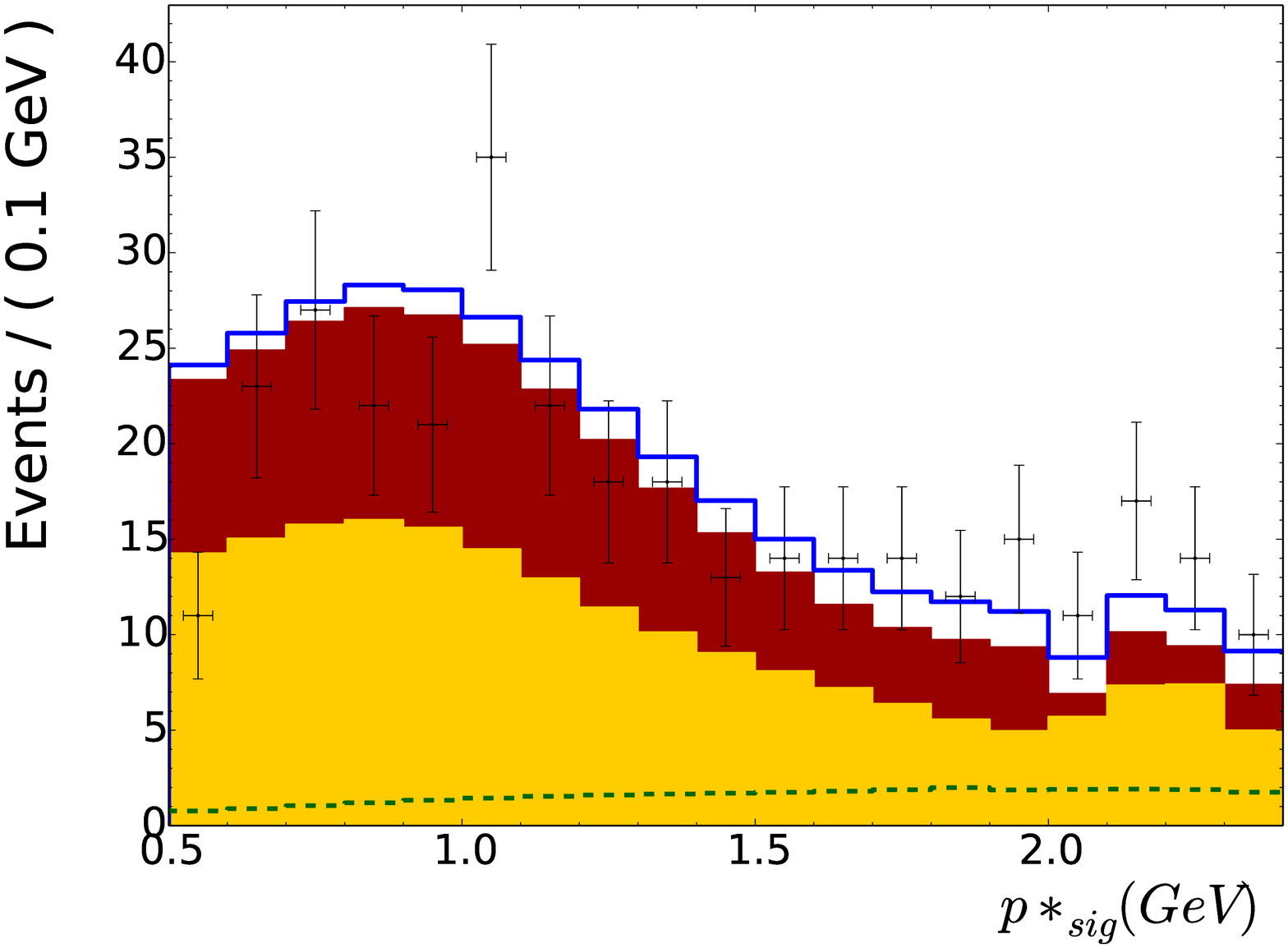} \put(-30,60){(d)}\\
\caption{Distribution of $\psig$, projected in the region $\ecl < 0.2$ GeV for (a)~$\tautomu$, (b)~$\tautoe$, (c)~$\tautopi$, and (d)~$\tautorho$. The markers show the data distribution, the solid line the total fitted distribution, and the dashed line the signal component. The orange (red) filled distribution represents the $B \bar{B}$ (continuum) background.}
\label{Fig:pcms}
 \end{figure}

To extract the number of reconstructed signal events, we perform an extended two-dimensional unbinned maximum-likelihood fit in $\psig$ and $\ecl$.
We use smoothed histogram probability density functions (PDFs)~\cite{smooth} obtained from MC to describe the signal and background components arising from events containing a $B\bar{B}$ pair. We use the product of one-dimensional PDFs for all components except for the signal in $\tautopi$ and $\tautorho$. In these modes the significant amount of cross-feed from other decay channels with additional, undetected neutral pions leads to a correlation between $\ecl$ and $\psig$ and the distributions are therefore described by two-dimensional histogram PDFs. The continuum background, including $e^+ e^- \to q \bar{q} \, (q = u, d, s, c)$, $\tau^+ \tau^-$, and two-photon events, is described using the off-resonance data and are scaled according to the relative luminosities. Since the off-resonance data sample is very limited, its $\ecl$ distribution is described by a linear function. The ratio of the normalisations of the  background components is fixed in the fit. We validate the distributions of the variables used in the fit and also others, e.g., $\bdl$, the outputs of the MVSs, and the missing energy in the event, by examining various control samples including the validation of the signal distribution in $\ecl$ and $\psig$ using the double-tagged sample, which reveals no significant discrepancy between data and MC.
The following five parameters are floated in the fit to the data to determine the signal branching fraction: ${\cal B}(\Bdecay)$ and the normalization of the background in each of the $\tau$ decay channels. The relative signal yields in the $\tau$ decay channels are constrained by the ratios of the reconstruction efficiencies. Figure~\ref{Fig:ecl} shows the $\ecl$ distribution and Fig.~\ref{Fig:pcms} shows the $\psig$ distribution projected in the region $\ecl < 0.2$ GeV.
We obtain a total signal yield of $N_{\rm sig} = 222 \pm 50$. This results in a branching fraction of ${\cal B}(\Bdecay) = (1.25 \pm 0.28) \times 10^{-4}$. The signal yields and branching fractions, obtained from fits for each of the $\tau$ decay modes separately are given in Table~\ref{tab:brSingle}.

 \begin{table}[th]
  \begin{tabular}{l @{\hspace{1em}} r@{$\pm$}l @{\hspace{1em}} r@{$\pm$}l}
  \hline \hline
   Decay Mode  & \multicolumn{2}{c}{$N_{\rm sig}$} & \multicolumn{2}{c}{${\cal B} (10^{-4})$} \\ \hline 
   $\tautomu$  & $13$  &$21$   & $0.34$  &$0.55$ \\
   $\tautoe$   & $47$  &$25$   & $0.90$  &$0.47$ \\
   $\tautopi$  & $57$  &$21$   & $1.82$  &$0.68$ \\
   $\tautorho$ & $119$ &$33$   & $2.16$  &$0.60$ \\
   Combined    & $222$ &$50$   & $1.25$  &$0.28$ \\
   \hline \hline
  \end{tabular}
  \caption{Signal yields and branching fractions, obtained from fits for the $\tau$ decay modes separately and combined.}
  \label{tab:brSingle}
 \end{table}

\par 
The list of systematic errors is given in Table~\ref{tab:sys}. The following systematic errors are determined by varying the corresponding parameters by their uncertainty, repeating the fit and taking the difference to the nominal fit result as systematic error: The normalization and slope of the continuum background component; the signal reconstruction efficiency; the branching fractions of dominant background decays, e.g. $B^- \to D^0 \ell^+ \nu_\ell$ followed by $D^0 \to K_L K_L$ or $D^0 \to K_L K_L K_L$; the correction of the tagging efficiency, obtained from the double tagged samples; and the branching fractions of the $\tau$ lepton.
To estimate the effect of the uncertainty on the shape of the histogram PDFs due to the statistical uncertainty in the MC data, the content of each bin is varied following a Poisson distribution with the original content as mean before the fit is performed. This is repeated 1000 times and the width of the distribution of branching fractions is taken as systematic error. For the systematic related to the best candidate selection, we perform the selection and the fit without applying the best candidate selection, thus allowing for multiple candidates per event. The result is divided by the average multiplicity of $1.07$ and compared to the nominal fit result. The uncertainty on the efficiency of the reconstruction of charged tracks and neutral pions and on the efficiency of the particle identification have been estimated using high statistics control samples. The charged track veto has been tested using the $D^0 \pi^+$ double-tagged sample by comparing the number of additional charged tracks in MC and data events. We find, that it agrees well, so we take the relative uncertainty on the number as systematic error. We also test an alternative description of the continuum background in $\ecl$ by using a polynomial of second order, but the deviation is well covered by the related systematic error, so we do not include it separately.
The quadratic sum of all contributions is $22.0 \%$. 

\begin{table}
\centering
\begin{tabular}{l r}
\hline \hline
Source & Relative Uncertainty (\%) \\ \hline
Histogram PDF shapes  				&  8.5  \hspace{3em} ${}$\\
Continuum description 				& 14.1  \hspace{3em} ${}$\\
Signal reconstruction efficiency 		&  0.6  \hspace{3em} ${}$\\
Background branching fractions 			&  3.1  \hspace{3em} ${}$\\
Efficiency calibration			 	& 12.6  \hspace{3em} ${}$\\
$\tau$ decay branching fractions  		&  0.2  \hspace{3em} ${}$\\
Best candidate selection			&  0.4  \hspace{3em} ${}$\\
Charged track reconstruction			&  0.4  \hspace{3em} ${}$\\
$\pi^0$ reconstruction 		 		&  1.1  \hspace{3em} ${}$\\
Particle identification 			&  0.5  \hspace{3em} ${}$\\
Charged track veto 				&  1.9  \hspace{3em} ${}$\\
Number of $B\bar{B}$ pairs			&  1.4  \hspace{3em} ${}$\\
Total & 22.0 \hspace{3em} ${}$\\
\hline \hline
\end{tabular}
\caption{List of systematic errors.}
\label{tab:sys}
\end{table}

\par
We exclude the hypotheses of no $\Bdecay$ decays with a significance of $3.8 \sigma$, by the convolution of the likelihood curve with a Gaussian distribution with a width of the systematic error. The significance is given by $\sqrt{2 \ln({\cal L} / {\cal L}_0) }$, where ${\cal L}_0$ is the likelihood of the hypotheses asuming zero signal events.

\par
In summary, we report the measurement of the branching fraction of $\Bdecay$ decays using a sample of $772 \times 10^6$ $B\bar{B}$ pairs, which we analyzed with the semileptonic tagging method. We measure it to be
\begin{equation*}
{\cal B}(\Bdecay) = [1.25 \pm 0.28 ({\rm stat.}) \pm 0.27({\rm syst.})] \times 10^{-4}
\end{equation*}
with a significance of $3.8 \sigma$. This result supersedes the previous measurement of the Belle collaboration~\cite{BelleLep}. It is consistent with previous measurements and with the SM expectation. We plan to combine this result with the recent measurement of the Belle collaboration using hadronic tagging~\cite{BelleHad} taking into account all relevant correlations of systematic errors.

\par
We thank the KEKB group for the excellent operation of the accelerator; the KEK cryogenics group for the efficient operation of the solenoid; and the KEK computer group, the National Institute of Informatics, and the PNNL/EMSL computing group for valuable computing and SINET4 network support.  We acknowledge support from the Ministry of Education, Culture, Sports, Science, and Technology (MEXT) of Japan, the Japan Society for the Promotion of Science (JSPS), and the Tau-Lepton Physics Research Center of Nagoya University; the Australian Research Council and the Australian Department of Industry, Innovation, Science and Research; the National Natural Science Foundation of China under contract No.~10575109, 10775142, 10875115 and 10825524; the Ministry of Education, Youth and Sports of the Czech Republic under contract No.~LA10033 and MSM0021620859; the Department of Science and Technology of India; the Istituto Nazionale di Fisica Nucleare of Italy; the BK21 and WCU program of the Ministry Education Science and Technology, National Research Foundation of Korea, and GSDC of the Korea Institute of Science and Technology Information; the Polish Ministry of Science and Higher Education; the Ministry of Education and Science of the Russian Federation and the Russian Federal Agency for Atomic Energy; the Slovenian Research Agency;  the Swiss National Science Foundation; the National Science Council and the Ministry of Education of Taiwan; and the U.S.\ Department of Energy and the National Science Foundation. This work is supported by a Grant-in-Aid from MEXT for Science Research in a Priority Area (``New Development of Flavor Physics''), and from JSPS for Creative Scientific Research (``Evolution of Tau-lepton Physics'').


\begin{thebibliography}{99}

\bibitem{CC}
Throughout this paper, the inclusion of the charge-conjugate decay mode is implied unless otherwise stated.

\bibitem{CKMfitter}
CKMfitter Group (J.~Charles {\it et al.}), Eur. Phys. J. C {\bf 41}, 1-131 (2005), updated result as of winter 2014 from http://ckmfitter.in2p3.fr.

\bibitem{BaBarLep}
B. Aubert {\it et al.} (BABAR Collaboration), Phys. Rev. D {\bf 81}, 051101 (2010).

\bibitem{BaBarHad}
J. P. Lees {\it et al.} (BABAR Collaboration), Phys. Rev. D {\bf 88}, 031102 (2013).

\bibitem{BelleLep}
K. Hara {\it et al.} (Belle Collaboration), Phys. Rev. D {\bf 82}, 071101 (2010).

\bibitem{BelleHad}
K. Hara {\it et al.} (Belle Collaboration), Phys. Rev. Lett. {\bf 110}, 131801 (2013).

\bibitem{PDG}
K. A. Olive {\it et al.} (Particle Data Group), Chin. Phys. C {\bf 38}, 090001 (2014).

\bibitem{kekb}
S. Kurokawa and E. Kikutani, Nucl. Instr. and Meth. A {\bf 499}, 1 (2003) and other papers included in this volume;
T.Abe et al., Prog. Theor. Exp. Phys. {\bf 2013}, 03A001 (2013) and following articles up to 03A011.

\bibitem{evtgen}
D.~J.~Lange, Nucl. Instr. and Meth. A {\bf 462}, 152 (2001).

\bibitem{GEANT}
R.~Brun {\it et al.}, GEANT, CERN Report No. DD/EE/84-1 (1984).

\bibitem{Belle}
A.~Abashian {\it et al.} (Belle Collaboration), Nucl. Instr. and Meth. A {\bf 479}, 117 (2002);
also see detector section in J.Brodzicka et al., Prog. Theor. Exp. Phys. {\bf 2012} 04D001 (2012).

\bibitem{svd2}
Z.Natkaniec {\it et al.} (Belle SVD2 Group), Nucl. Instr. and Meth. A {\bf 560}, 1(2006);
Y. Ushiroda (Belle SVD2 Group), Nucl. Instr. and Meth.A {\bf 511} 6 (2003). 

\bibitem{NB}
M.~Feindt and O. Kerzel, Nucl. Instr. and Meth. A {\bf 559}, 190-194 (2006).

\bibitem{HadFR}
M.~Feindt {\it et al.}, Nucl. Instr. and Meth. A {\bf 654}, 432-440 (2011).

\bibitem{FWM}
S. H. Lee {\it et al.} (Belle Collaboration), Phys. Rev. Lett. {\bf 91}, 261801 (2003).

\bibitem{CleoCones}
D. M. Asner {\it et al.} (CLEO Collaboration), Phys. Rev. D {\bf 53}, 1039 (1996).

\bibitem{smooth}
V. Blobel, ``Smoothing of Poisson distributed data.'' http://www.desy.de/∼blobel/splft.f .

\end{thebibliography}
\end{document}